\documentclass[authoryear]{FLO_v1}%

\usepackage{graphicx}
\usepackage{upgreek}
\usepackage{multicol,multirow}
\usepackage{amsmath,amssymb,amsfonts}
\usepackage{mathrsfs}
\usepackage{amsthm}
\usepackage[figuresright]{rotating}
\usepackage{appendix}
\usepackage[authoryear]{natbib}
\usepackage{ifpdf}
\usepackage[T1]{fontenc}
\usepackage{newtxtext}
\usepackage{newtxmath}
\usepackage{textcomp}
\usepackage{xcolor}
\usepackage{natbib}
\usepackage{xfrac}
\usepackage{tikz}
\usetikzlibrary{arrows.meta,patterns,decorations.pathreplacing}
\usepackage{booktabs}
\usepackage{overpic}

\usepackage[colorlinks,allcolors=blue]{hyperref}
\definecolor{jourcolor}{cmyk}{1,0.57,0.01,0.38}
\hypersetup{
    colorlinks,%
    citecolor=jourcolor,%
    filecolor=jourcolor,%
    linkcolor=jourcolor,%
    urlcolor=jourcolor
}

\theoremstyle{definition}

\def\UH{U_\mathrm{H}}

\newcommand{\pt}[1]{\frac{\partial {#1}}{\partial t}}
\newcommand{\px}[1]{\frac{\partial {#1}}{\partial x}}
\newcommand{\pz}[1]{\frac{\partial {#1}}{\partial z}}
\newcommand{\pxx}[1]{\frac{\partial^2 {#1}}{\partial x^2}}
\newcommand{\pzz}[1]{\frac{\partial^2 {#1}}{\partial z^2}}
\def\e{\mathrm{e}}
\def\sgn{\text{sgn}}

\makeatletter
\def\ps@titlepage{\let\@oddhead\@empty\let\@evenhead\@empty
  \let\@oddfoot\@empty\let\@evenfoot\@empty}
\def\ps@headings{\def\@oddfoot{\hfill}\let\@evenfoot\@oddfoot
  \def\@evenhead{{\headerfont\thepage\qquad\textit{\rightmark}\hfill}}%
  \def\@oddhead{{\headerfont\hfill\textit{\leftmark}\qquad\thepage}}%
  \let\@mkboth\markboth}
\makeatother
\articletype{RESEARCH ARTICLE}

\begin{document}

\title[Two-way coupling of gravity waves and wind farm wakes: a reduced-order numerical model]{Two-way coupling of gravity waves and wind farm wakes: a reduced-order numerical model}

\author[Hossein A. Kafiabad and Majid Bastankhah]{Hossein A. Kafiabad$^{1\ast}${{\href{https://orcid.org/0000-0002-8791-9217}{\includegraphics{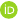}}}} and Majid Bastankhah$^{2\ast}${{\href{https://orcid.org/0000-0002-1326-027X}{\includegraphics{orcid_logo}}}}}

\address[1]{Department of Mathematical Sciences, Durham University, Durham DH1 3LE, UK}
\address[2]{Department of Engineering, Durham University, Durham DH1 3LE, UK}

\corres{*}{Corresponding author. E-mail:
\emaillink{hossein.amini-kafiabad@durham.ac.uk,majid.bastankhah@durham.ac.uk}}

\keywords{wind farm wakes; gravity waves, reduced-order numerical model}

\date{Accepted for publication in \textit{Flow}}

\abstract{This paper develops a reduced-order numerical framework for modelling the two-way coupling between gravity waves and turbulent wakes in large-scale wind farms. Starting with the non-hydrostatic Boussinesq equations and introducing simplifications appropriate to the boundary layer and the overlying stratified free atmosphere yield separate governing equations for the two regions. These are coupled through a dynamic boundary condition at the capping inversion, which directly captures the feedback of gravity waves on the boundary-layer flow. A mixed pseudo-spectral finite-difference discretisation yields a
computationally efficient model while retaining vertical boundary-layer
structure. For the semi-infinite wind farm considered here, a
complete solution is obtained in a matter of seconds on a standard laptop. Comparisons with large-eddy simulations (LES) confirm the model successfully reproduces both internal wind-farm flow and large-scale gravity-wave effects. It captures the upstream blockage induced by adverse pressure gradients, as well as the accelerated wake recovery within and downwind of the farm, driven by favourable pressure gradients.}

\maketitle

\begin{boxtext}

\textbf{\mathversion{bold}Impact Statement}

The continued growth in the size and hub height of modern wind turbines has led
to wind farms that occupy a substantial fraction of the atmospheric boundary
layer. The reduction in wind speed across such farms displaces the capping
inversion and excites gravity waves in the stratified free atmosphere aloft,
generating pressure gradients that modify the flow within the farm. Wake
evolution and gravity-wave dynamics therefore constitute a single coupled
problem. The present work addresses this coupling by retaining the vertical
structure of the boundary layer together with the non-hydrostatic dynamics of the
free atmosphere, joined through a dynamic condition at the inversion.

The developed numerical framework provides a mid-fidelity tool for bridging the gap between computationally expensive high-fidelity simulations and highly parametrised engineering models of wind-farm--atmosphere interaction. The present paper demonstrates this capability for a semi-infinite wind farm, where the model reproduces the LES-predicted blockage and wake recovery at a computational cost of seconds. Once extended to finite farms and validated over a wider range of conditions, the combination of computational efficiency with a physics-based representation of wake and gravity-wave coupling could make the framework a useful tool for parameter sweeps, for assessing farm--farm interactions, and potentially for informing wind-farm parametrisations in mesoscale models.

\end{boxtext}

\section{Introduction}

\begin{figure}
    \centering
    \includegraphics[width=.95\linewidth]{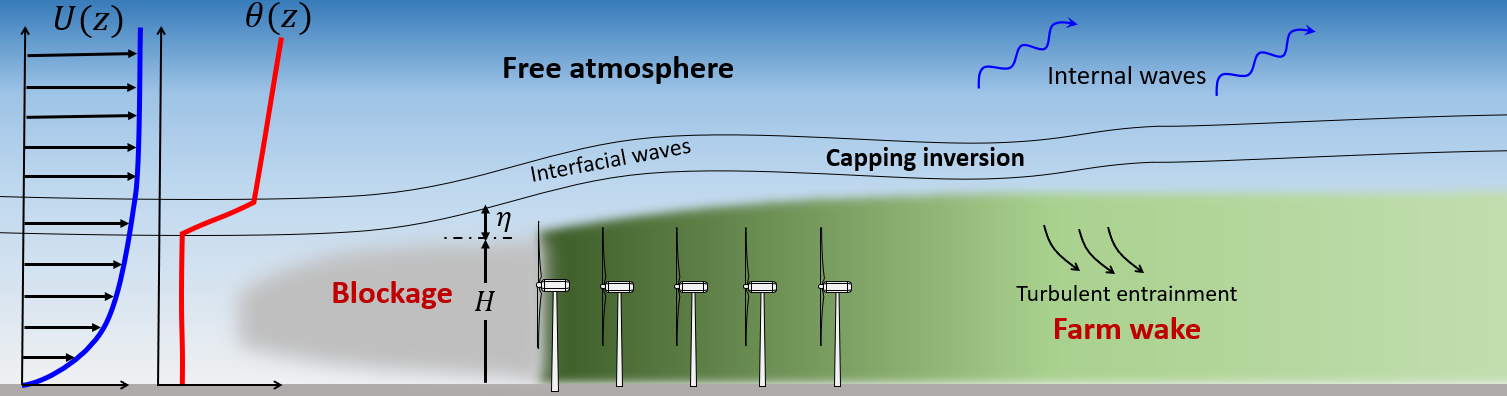}
    \caption{Interaction of a wind farm with a CNBL. The red and blue curves show schematic potential-temperature and background velocity profiles, respectively.}
    \label{fig:schematic}
\end{figure}

The rapid growth of the wind energy sector has brought wind farms to a scale that is no longer small relative to the surrounding atmosphere. While older farms could be modelled as modest perturbations to the Earth's surface roughness, modern utility-scale farms are now comparable in size to the atmospheric boundary layer (ABL) thickness. This increase in scale creates a complex two-way interaction between the wind farm and the atmosphere, making phenomena that were negligible for smaller farms central to understanding both wind-farm performance and atmospheric-flow response. A key driver of these phenomena is the thermal stratification within the atmosphere. Even when the boundary layer itself is neutral, stratification is typically present above the ABL in the form of an abrupt temperature jump at the capping inversion and a more gradual variation in the free atmosphere above (figure \ref{fig:schematic}). This structure is called a conventionally neutral boundary layer (CNBL).  In this case, the wind farm's interaction with the overlying stratified layers generates large-scale gravity waves. Specifically, the extraction of energy by turbines and the subsequent reduction of wind speed at hub height create vertical motions in the ABL to satisfy continuity. These vertical motions displace the capping inversion, generating \textit{interfacial waves}, associated with oscillations of the capping inversion, and \textit{internal waves}, supported by stratification in the free atmosphere. The generation of these gravity waves induces pressure gradients in the ABL that feed back on the wind speed at hub height \citep{allaerts_gravity_2018,allaerts2019sensitivity}. Beyond causing significant wind-farm blockage \citep{lanzilao_parametric_2024}, this  pressure gradient can impact the recovery of turbulent wakes within and downwind of wind farms \citep{lanzilao_wind-farm_2025}. %

The high-fidelity approach for modelling gravity waves and farm turbulent wakes is Large Eddy Simulation (LES) \citep{allaerts2017,allaerts_gravity_2018,maas_gigawatt_2023,lanzilao_parametric_2024,lanzilao_wind-farm_2025,stipa2024tosca,khan_investigating_2025,khan_dependence_2026}. However, properly capturing gravity waves in LES is computationally expensive and challenging as the massive spatial extent of gravity wave propagation requires significantly large domains. Moreover, the sensitivity of these waves to boundary conditions demands careful numerical treatment \citep{lanzilao_parametric_2024}. These expensive simulations might be therefore unsuitable for large parameter sweeps, layout optimisation, or real-time farm control; precisely the areas where the wind-energy industry most urgently needs reliable tools.

To tackle these challenges, earlier studies developed reduced-order models to represent gravity-wave effects in wind farms. \cite{smith_gravity_2010} introduced the foundational model by linearising the ABL governing equations for vertically uniform perturbation fields and relating the pressure perturbation to capping-inversion displacement through internal and interfacial wave dynamics. While successful in qualitatively capturing the impact of gravity waves, this approach relied on two simplifying assumptions: the flow fields and farm forcing were vertically uniform, and the forcing was prescribed independently of the wave response. Subsequent studies tried to address these limitations. \cite{allaerts2019sensitivity} introduced a multi-layer approach, splitting the boundary layer into a farm layer (with farm forcing) and an upper layer (without farm forcing). Other studies \citep[e.g.,][]{stipa_multi-scale_2024,devesse_mesomicro_2024} further developed this approach by coupling linearised gravity-wave models with engineering wake models through an iterative procedure to capture the two-way interaction between wake evolution and gravity waves. Despite this important progress, %
the multi-layer approach does not fully capture the continuous vertical variation of the boundary-layer flow. Second, existing approaches couple linearised gravity-wave models with engineering wake models, resulting in a highly parametrised system. The wave component typically requires parametrisations of turbulent diffusion and interfacial shear stresses, while engineering wake modelling introduces additional choices such as the single wake model, wake recovery rate, superposition method, and local blockage model \citep{devesse_mesomicro_2024}. %

A separate line of work has instead used reduced-order numerical models, and in particular linearised CFD approaches such as FUGA \citep{ott2011Fuga} and other works  \citep[e.g.,][]{ebenhoch2017,segalini_linearized_2017,segalini_analytical_2021}. In this approach, the turbine forcing can enter in governing equations as a momentum sink acting on a flow field, so that wakes and local blockage develop as part of the solution rather than being prescribed through an engineering wake model.  As a mid-fidelity tool, these models are expected to reproduce farm-scale flow physics more faithfully than engineering wake models, at a computational cost that still remains orders of magnitude below that of LES.  To date, however, these models have been applied mainly to intra-farm flows \citep{ebenhoch2017}, and more importantly are formulated for a boundary layer that is unbounded or capped by a rigid lid. Therefore, they cannot capture farm-scale blockage, which arises from the pressure gradients induced by gravity waves as the capping inversion is displaced.

To address these limitations, this work develops a new reduced-order numerical framework that resolves the continuous vertical structure of the ABL while capturing the two-way coupling between wake flows and gravity waves within a single set of governing equations. The flow inside the ABL is obtained from a
simplified numerical model in which the turbine forcing acts directly on the resolved flow field, so that no engineering wake model and iterative coupling between
micro- and meso-scale components are needed. The feedback of gravity waves on the ABL flow is captured through the boundary condition imposed at the top of the ABL. Here, for simplicity we focus on a semi-infinite wind
farm (finite in the streamwise direction and infinite in the lateral direction), but the developed numerical approach can be readily extended to finite wind farms in future works. %

\section{Model development}

\subsection{Governing equations}\label{sec:governing_equation}

We consider a background flow with horizontal velocity that only varies with height $U(z)$ and no vertical velocity. %
The steady two-dimensional non-hydrostatic Boussinesq equations yield
\begin{subequations}\label{2DBoussinesq}
\begin{align}
     U \px{u} + u \px{u} + w U' + w \pz{u} &= - \px{p} + \pz{\tau} + f, \label{BLlin_xmom}\\
    U \px{w} + u \px{w} + w \pz{w} &= - \pz{p} + b + \px{\tau}, \label{BLlin_zmom}\\
    \px{u} + \pz{w} &= 0\label{BLeq_linear_continuity}, \\
    U \px{b} + N^2 w &= 0,
\end{align}
\end{subequations}
where $u$ and $w$ are the perturbation horizontal and vertical velocities induced by the wind farm, respectilvey, and $U'=\mathrm{d}U/\mathrm{d}z$. The buoyancy perturbation is
$b = g\theta/\theta_0$ or $b = -g\rho/\rho_0$, where $\theta$ and $\rho$ denote potential temperature and density perturbations, respectively, and $\theta_0$ and $\rho_0$ are the corresponding reference values. The buoyancy frequency is
$N = \sqrt{-(g/\rho_0)\, d\bar{\rho}/dz} = \sqrt{(g/\theta_0)\, d\bar{\theta}/dz}$, where the overbar denotes background quantities. The forcing $f$ represents the horizontal force exerted by the wind farm (discussed later in \S\ref{sec:farm_forcing_shorter}), $\tau$ is the perturbation Reynolds shear stress, and $p$ is the modified perturbation pressure, including the pressure and the isotropic part of the Reynolds stress tensor (normal stresses). The pressure and turbulent stresses are scaled by $\rho_0$. We assume that \eqref{2DBoussinesq} holds in both the ABL and the lower free atmosphere. %
The presence of a wind farm leads to a displacement of the inversion layer denoted by $\eta$ (shown schematically in figure \ref{fig:schematic}). Considering small $\eta$, we approximate the ABL with a rectangular domain capped at $z=H$. However, we take the dynamical effects of this displacement into account leading to the coupling of the ABL and free atmosphere (discussed later in \S \ref{sec:coupling_ABL_FA}). We denote the perturbation values right above and below the inversion layer by superscripts $+$ and $-$, respectively. For example, $w^+$ and $w^-$ represent the vertical velocity immediately above and below the inversion.

Unlike the hydrostatic model of \cite{smith_gravity_2010}, where the vertical momentum equation is neglected throughout the atmospheric boundary layer (ABL) and the free atmosphere, we retain the full non-hydrostatic momentum equations in both regions. This differs also from subsequent studies \citep[e.g.][]{allaerts2019sensitivity,devesse_mesomicro_2024}, which incorporated non-hydrostatic effects through the free-atmospheric wave dispersion relation while retaining the vertically-averaged hydrostatic description of the ABL. Consequently, in the present formulation the horizontal velocity, vertical velocity and pressure remain dynamically coupled throughout both regions, making the vertical acceleration an integral part of the model.%

It is also worth noting that the non-linear advection terms are retained in \eqref{2DBoussinesq}. Previous reduced-order numerical works \citep[e.g.][]{ott2011Fuga,ebenhoch2017} usually linearise the governing equations by dropping all products of perturbation quantities. Such a simplification is formally
valid only where the perturbations remain small compared with the background flow, which may not be necessarily the case near the turbines. The turbine forcing is also non-linear for the same reason, since the thrust force scales with $(U+u)^2$. We therefore retain the non-linear advection and forcing terms when solving for the flow within the boundary layer (\S\ref{sec:boundary_layer}).
In the free atmosphere (\S\ref{sec:free_atm}), however, the flow is not directly forced by the turbines, so perturbation amplitudes remain small relative to the background wind. The non-linear advection terms in \eqref{2DBoussinesq} are therefore expected to be small in this region, or at least they become considerable only at scales far larger than those of the  wind-farm-induced gravity waves \citep{vallis2017atmospheric}. 

\subsection{Boundary layer}\label{sec:boundary_layer}
Within the boundary layer, we assume constant density, so $N=b=0$. Rearranging the remaining terms in \eqref{2DBoussinesq} and eliminating $u$ and $p$ gives the following equation for $w$:
\begin{equation}\label{BLeq4w_tau}
   U \nabla^2 w - U'' w +\mathcal{N}(w) = -\pzz{\tau}+\pxx{\tau}- \pz{f}.
\end{equation}
where $\nabla^2 = \tfrac{\partial^2}{\partial z^2}+\tfrac{\partial^2}{\partial x^2} $ and $\mathcal{N}(w)$ represents the nonlinear terms:
\begin{equation}\label{eq:non-linear}
    \mathcal{N}(w) =  - \nabla^2 w \int_0^x \frac{\partial w }{\partial z}(s,z) \, \mathrm{d}s  + w\left(\frac{\partial^2 w}{\partial x \partial z} + \int_0^x \frac{\partial^3 w }{\partial z^3}(s,z) \, \mathrm{d}s  \, \right),
\end{equation}
with $s$ being a dummy variable for integration. The shear stress is modelled using the Boussinesq turbulent-viscosity hypothesis
with a mixing-length closure \citep{belcher-Hunt2003}, applied to the total
velocity field,
\begin{equation}\label{eq:tau_total}
    \tau_{\mathrm{tot}} = l_m^2 \left| \pz{(U+u)} \right| \pz{(U+u)},
\end{equation}
where $l_m(z)=\kappa z$ is the mixing length and $\kappa=0.4$ is the von K\'arm\'an
constant. \cite{belcher-Hunt2003}, who developed a model for the adjustment of a boundary layer to a canopy of roughness elements, modified the mixing length within dense canopies to account for the eddies generated in the shear layer at the canopy top. Since a wind farm can be regarded as a sparse canopy, we instead assume that the structure of the turbulence in the approaching boundary layer is not substantially altered by the farm and retain $l_m=\kappa z$ throughout. An important limitation of this assumption is that it ignores the effect of the turbulence added by the turbines on the turbulent viscosity. Assuming the total shear remains positive, so that the modulus may be
dropped, and expanding gives
\begin{equation}\label{eq:tau_expanded}
    \tau_{\mathrm{tot}} = l_m^2 \left( U'^2 + 2U' \pz{u} +
    \left( \pz{u} \right)^{2} \right),
\end{equation}
in which the first term is the background stress carried by the undisturbed
profile. Subtracting it and neglecting the quadratic term, which is small provided the
perturbation shear is small compared with the background shear, leaves the
perturbation stress
\begin{equation}\label{eq:tau_modelled}
    \tau = 2\nu_t(z) \pz{u},
\end{equation}
where $\nu_t = l_m^2 \, \mathrm{d}U/\mathrm{d}z$ is the turbulent viscosity of the background profile; $U(z)$ can be any prescribed profile, and the one used for the cases presented in this paper is given in \S\ref{sec:result}. The factor of two therefore arises from linearising the quadratic closure, and $\nu_t$ depends on the background shear alone, so the farm's contribution to the mixing is not represented. This linearisation is the one place where perturbations are assumed small within the boundary layer, in
contrast to the advection and forcing terms, which are retained in full. The two cases are not equivalent, however: the advection and forcing terms appear exactly in the governing equations, whereas \eqref{eq:tau_total} is itself a model of the unresolved turbulence that remains approximate regardless of the method we choose. Retaining its full form would arguably be more realistic, since it lets the eddy viscosity
respond to the perturbed shear, but it makes $\tau$ quadratic in the
perturbation and complicates the solution procedure. We therefore adopt the linear form, while noting that it may under-represent the enhanced mixing within dense farms.

Substituting \eqref{eq:tau_modelled} into \eqref{BLeq4w_tau} and using the continuity equation \eqref{BLeq_linear_continuity} yields
\begin{equation}\label{realBLeq4w}
   U \nabla^2 w  - U'' w -2\pzz{}\left( \nu_t(z) \pzz{} \int_0^x w(s,z) \, \mathrm{d}s \right) +2\nu_t(z)\px{}\pzz{w} + \mathcal{N}(w) = - \pz{f},
\end{equation}
In the streamwise direction, we impose periodic boundary conditions, which are natural to the spectral discretisation introduced in \S\ref{sec:numerics}. Strictly, a periodic solution describes an infinite streamwise array of identical farms separated by the domain length $L_x$ rather than a single farm. It approximates an isolated farm only when $L_x$ is large enough for the perturbations induced by the neighbouring images to be negligible over the region of interest; the domain length required for this is established in \S\ref{sec:result}. Four boundary conditions are required in the vertical to ensure the closure of \eqref{realBLeq4w}. The lower boundary of the computational domain is placed at the aerodynamic roughness length $z=z_0$, i.e.\ the height at which the logarithmic background profile vanishes, $U(z_0)=0$, where the no-slip condition and \eqref{BLeq_linear_continuity} imply $w = 0$ and $\tfrac{\partial w}{\partial z} = 0$. At the top of the ABL, we neglect $\mathcal{N}(w)$, $\tau$, and $f$ in \eqref{BLlin_xmom}, since the turbine forcing is assumed to be confined sufficiently below the capping inversion and turbulent and nonlinear effects diminish with height. This yields $ \nabla^2 w - (U''/U) w = 0$ at $z=H$. This assumption restricts the present model to farms whose turbines do not extend into the immediate vicinity of the inversion layer. The final other boundary condition at the top of the ABL is obtained by coupling with the free atmosphere which will be discussed in section \S \ref{sec:coupling_ABL_FA}.

\subsection{Wind farm forcing}\label{sec:farm_forcing_shorter}

The governing equations of \S\ref{sec:boundary_layer} describe a two-dimensional
flow in the $(x,z)$ plane, whereas the flow through a wind farm is inherently
three-dimensional. The two are connected by a lateral average, which we
introduce here since it also determines the form of the forcing term $f$
appearing in \eqref{BLlin_xmom}. For a farm that is infinite in the lateral
direction with a spanwise turbine spacing $S_y$, the flow is periodic in $y$
with period $S_y$, and we define the lateral average of a field $\phi$ as
\begin{equation}\label{eq:lat_avg_def}
    \langle \phi \rangle (x,z) = \frac{1}{S_y}
    \int_{-S_y/2}^{\,S_y/2} \phi(x,y,z)\, \mathrm{d}y ,
\end{equation}
and write $\phi = \langle \phi \rangle + \phi^{\mathrm{d}}$, where $\phi^{\mathrm{d}}$ denotes the
departure from the lateral mean. Applying \eqref{eq:lat_avg_def} to the
three-dimensional Boussinesq equations, the terms in $\partial/\partial y$
vanish by periodicity, while the advection terms generate dispersive stresses
such as $\partial_x\langle u^{\mathrm{d}}u^{\mathrm{d}}\rangle$ and $\partial_z\langle u^{\mathrm{d}}w^{\mathrm{d}}\rangle$.
These are neglected here for simplicity, so that \eqref{2DBoussinesq} is
recovered for the laterally averaged fields, although previous studies
\citep[e.g.,][]{bastankhah2024farmwake,devesse_mesomicro_2024} indicate that
their impact can be non-negligible for strongly heterogeneous intra-farm flows.
The lateral heterogeneity of the flow is nonetheless retained in the turbine
forcing introduced later in this section.
\S\ref{sec:governing_equation}--\S\ref{sec:boundary_layer} are therefore
understood as laterally averaged quantities, with the angle brackets omitted for
brevity in other parts of this paper, and the forcing $f$ is the lateral average of the three-dimensional turbine force derived below.

Each turbine is represented as an actuator disc of zero streamwise thickness,
uniformly loaded over its frontal area, so that rotation and the radial
variation of the loading are not represented. For a semi-infinite wind farm with
$N_r$ rows of turbines, let $n\in\{1,2,\dots,N_r\}$ denote the row number
increasing in the streamwise direction, $x_n$ the streamwise location of the
$n^{\text{th}}$ row, $z_h$ the hub height, and $y_n + q S_y$
with the integer $q$ labelling turbines within the row. The
total thrust of a single turbine of the $n^{\text{th}}$ row is
$T_n = \tfrac{1}{2}\rho_0 C'_{T,n} U_n^2 \pi R^2$, where $U_n$ is the local
disc-averaged velocity of WT$_n$s (i.e., turbines in the $n^{th}$ row),
$C'_{T,n}=4C_{T,n}/(1+\sqrt{1-C_{T,n}})^2$ is the local thrust coefficient of
the WT$_n$s, and $R=D/2$ is the turbine radius. Distributing this force
uniformly over the disc area $\pi R^2$ gives the three-dimensional force per
unit mass exerted by the farm,
\begin{equation}\label{eq:turbine_forcing_3D}
    \mathcal{F}(x,y,z) = -\frac{1}{2}\sum_{n=1}^{N_r}
    C'_{T,n}\, U_n^2\, \delta(x-x_n)
    \sum_{q=-\infty}^{\infty} \mathcal{H}\!\left(R - r_{n,q}\right),
    \qquad
    r_{n,q} = \sqrt{(y - y_n - q S_y)^2 + (z-z_h)^2},
\end{equation}
where $\delta(x)$ is the Dirac delta function, $\mathcal{H}$ is the Heaviside
step function, and $r_{n,q}$ is the radial distance from the centre of the
corresponding rotor. %
By averaging \eqref{eq:turbine_forcing_3D} laterally based on \eqref{eq:lat_avg_def}, %
 the laterally averaged force per unit mass exerted by the farm $f$ becomes
\begin{equation} \label{eqs:turbine_forcing}
f = \langle \mathcal{F} \rangle =
\begin{cases}
-\dfrac{1}{S_y}\displaystyle\sum_{n=1}^{N_r} C'_{T,n}\, U_{n}^2\, \delta(x-x_n)\sqrt{R^2-(z-z_h)^2}, & |z - z_h| < R,\\[8pt]
0, & |z - z_h| \ge R,
\end{cases}
\end{equation}
The vertical profile of forcing in \eqref{eqs:turbine_forcing} is therefore simply the
half-chord of the rotor disc scaled by the lateral inter-turbine spacing, i.e. the local
solidity of the turbine row: the force is largest at hub height and tapers to
zero at the rotor tips. For numerical implementation, the forcing term
is smoothed in both the streamwise and vertical directions using a Gaussian kernel. The Dirac delta in \eqref{eqs:turbine_forcing} is replaced by a Gaussian of standard deviation $\sigma_x=\Delta x$, where $\Delta x$ is the streamwise grid spacing (between $0.2D$ and $D$ for the cases presented in \S\ref{sec:result}), and the vertical profile is convolved with a Gaussian of standard deviation $\sigma_z=D/20$, which regularises the square-root behaviour of the half-chord at the rotor tips. The results are not highly sensitive to these choices. Halving or doubling $\sigma_x$ changes the maximum hub-height velocity deficit by less than 4\% for the cases of \S\ref{sec:result}.

Since the governing equations are solved for the laterally averaged perturbation
field $u(x,z)$, they carry by construction no information about the lateral
variation of the flow within the farm. The thrust force in
\eqref{eqs:turbine_forcing}, however, depends on the velocity actually incident
on the rotors, which differs from its laterally averaged counterpart. For example, in an
aligned farm the turbines operate in the wakes of the upstream rows and
therefore experience a deficit considerably larger than the laterally averaged
one, whereas in a staggered farm the two (i.e., the local and laterally averaged
deficits) are comparable after the first a few rows. Without a relation
linking the two, the model would incorrectly return identical predictions for the two
layouts. We therefore write
\begin{equation}\label{eq:Un_beta}
U_n = U_h + \beta_n\, u(x_n,z_h),
\end{equation}
where $U_h=U(z_h)$ and $\beta_n$ is the ratio between the local wake deficit
induced by upstream turbines and its laterally averaged value at the same
streamwise location. We estimate $\beta_n$ using a simplified version of the
method developed by \cite{bastankhah2024farmwake}.
Denoting by $u_j$ the perturbation velocity induced by WT$_j$s and by $\langle
u_j\rangle$ its lateral average, linear superposition of wake effects gives
\begin{equation}\label{eq:beta_super}
\beta_n=\frac{\displaystyle\sum_{j=1}^{n-1} u_{j}(x_n,y_n,z_h)}
{\displaystyle\sum_{j=1}^{n-1} \langle u_{j}\rangle(x_n)} .
\end{equation}
Prescribing the deficit of each individual turbine wake as a Gaussian profile whose lateral width grows linearly with downstream distance \citep{bastankhah2014new}, and summing over the infinite lateral column of WT$_j$ turbines provides
\begin{equation}\label{eq:theta_link}
u_{j}(x_n,y_n,z_h)=\langle u_{j}\rangle(x_n)\,
\vartheta_{3}\!\left[\frac{\pi(y_n-y_j)}{S_y},\,
\exp\!\left(-\frac{X_{jn}^{2}}{80\,(S_y/D)^{2}}\right)\right],
\qquad
X_{jn}=c_k\,\frac{x_n-x_j}{D}+10,
\end{equation}
where $\vartheta_3[\zeta,q]=1+2\sum_{l=1}^{\infty}q^{l^2}\cos(2l\zeta)$ is the Jacobi theta function. The constant $c_k$ is a tuning parameter and a value of $c_k=1.75$ provides acceptable predictions against the three LES datasets presented in \S\ref{sec:result}. Since the same datasets are used to select $c_k$ and to assess the model, the comparisons of \S\ref{sec:result} should be regarded as partly a calibration of this constant. The three cases nevertheless span a reasonable range of turbine spacings ($S_x/D=5$--$7.5$), roughness lengths and atmospheric conditions, and it is encouraging that a single value of $c_k$ provides good agreement across all three. Whether the same value remains appropriate outside this range (e.g. for substantially different thrust coefficients or farm layouts) has yet to be established. We also stress that $c_k$ only affects the layout correction $\beta_n$. It does not enter the coupling between the boundary layer and the gravity waves, which is the main subject of this paper, and it is inactive for staggered layouts, for which $\beta_n\approx1$ (see below). Equation \eqref{eq:theta_link} shows that the theta function expresses the local-to-laterally-averaged ratio of a single turbine row. Since $\vartheta_3[\zeta\pm\pi,q]=\vartheta_3[\zeta,q]$, the lateral coordinates $y_n$ and $y_j$ may denote the position of any turbine within the respective rows.

In contrast to \cite{bastankhah2024farmwake}, where the row contributions
$\langle u_j\rangle$ were taken from the evolving solution of the momentum
equations, here we seek a simple algebraic expression for $\beta_n$ that depends only
on the farm layout, so that $\beta_n$ can be evaluated before the flow
solution is obtained. To this end, the deficit amplitude is evaluated using the
far-wake asymptotic form of the Gaussian wake model \citep{bastankhah2014new},
in which the amplitude becomes linearly proportional to $C_{T,j}$. After algebraic
manipulations, this gives
\begin{equation}\label{eq:u_j_averaged}
    \langle u_{j}\rangle(x_n)/U_h = -2.5\sqrt{2\pi}\,C_{T,j}X_{jn}^{-1}(S_y/D)^{-1}.
\end{equation}
Combining \eqref{eq:beta_super}, \eqref{eq:theta_link} and
\eqref{eq:u_j_averaged}, and assuming that the thrust coefficients of turbines
within the farm are of comparable magnitude, yields
\begin{equation}\label{eq:beta_final}
\beta_n =
\left(
\displaystyle\sum_{j=1}^{n-1}
X_{jn}^{-1}\,
\vartheta_{3}\!\left[
\frac{\pi (y_n - y_j)}{S_y},\,
\exp\!\left(-\frac{X_{jn}^2}{80\,(S_y/D)^2}\right)
\right]
\right)
\Big/
\displaystyle\sum_{j=1}^{n-1} X_{jn}^{-1}\, .
\end{equation}
According to \eqref{eq:beta_final}, $\beta_n$ is determined solely by the
relative positions of the upstream rows: the lateral offset $y_n-y_j$ sets how
much of each upstream wake is felt at the rotor, while the weight $X_{jn}^{-1}$
ensures that nearer rows contribute more strongly. %

It is important to note a crude simplification when we applied $\beta_n$ in \eqref{eq:Un_beta} to the full perturbation
velocity. In reality, only the component of velocity perturbation induced by upstream wakes requires this correction. The flow perturbation associated with local blockage, or with the global blockage induced by gravity waves do not need this correction. We therefore clamp $\beta_n$ to a minimum of
one, so that these contributions are not spuriously reduced, and the first
row, unaffected by upstream wakes, likewise takes $\beta_1=1$. In practice, the correction therefore acts mainly in aligned layouts, where the local velocity is considerably lower than its laterally averaged value and \eqref{eq:beta_final} substantially exceeds unity. In staggered layouts, by contrast, the flow becomes comparatively homogeneous in the lateral direction after the first few rows, so that the two velocities are
close and no correction is applied.

\begin{figure}
    \centering
    \includegraphics[width=0.95\linewidth]{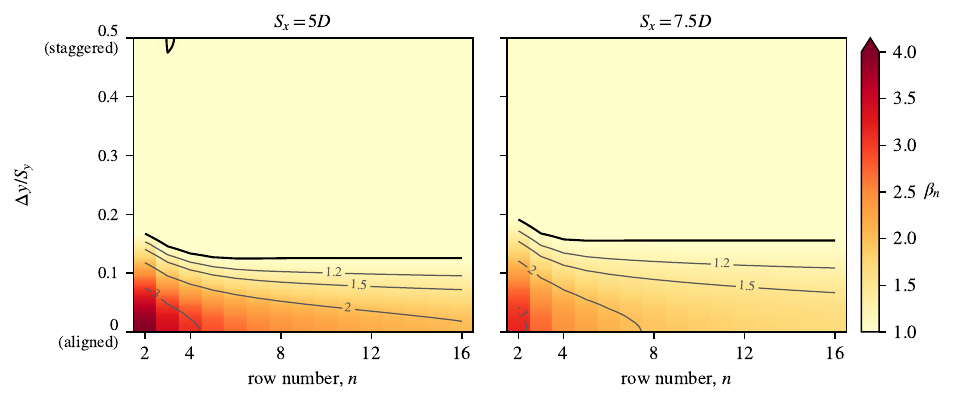}
    \caption{Variation of $\beta_n$ with row number and the lateral offset
    $\Delta y$ between consecutive rows, for $S_y=5D$ and two streamwise
    spacings. Zero offset corresponds to an aligned farm and $\Delta y/S_y=0.5$
    to a staggered one. The thick line marks $\beta_n=1$, below which the
    correction for lateral heterogeneity is active. The first row, for which
    $\beta_1=1$, is not shown.}
    \label{fig:beta_offset}
\end{figure}

Figure~\ref{fig:beta_offset} shows the variation of $\beta_n$
for farms in which each row is displaced laterally by a constant offset
$\Delta y$ relative to the row immediately upstream, so that
$y_n-y_j=(n-j)\Delta y$. Aligned and staggered farms correspond to the limiting
cases $\Delta y/S_y=0$ and $\Delta y/S_y=0.5$, respectively, and by the
periodicity of $\vartheta_3$ these two limits bracket all possible layouts of
this type. The correction is largest immediately downstream of the first row,
where the wakes are narrowest relative to the lateral spacing, reaching
$\beta_2\approx 4$ for an aligned farm with $S_x=5D$, and decays further downstream as the flow within the farm becomes more homogenous. The value of $\beta$ also reduces with increasing streamwise spacing, since
the wakes reaching a given row have then travelled farther. The
correction is confined to a comparatively narrow band of near-aligned layouts,
$\Delta y/S_y\lesssim 0.13$ for $S_x=5D$.

Finally, we note that the parameter $\beta_n$ is a consequence of solving for
the laterally averaged flow, and is therefore specific to the two-dimensional
formulation adopted here. It is not a simplification associated with the coupling between wakes and gravity waves, which is the main contribution of the present framework. In a three-dimensional extension, where the lateral variation of the flow is resolved, the turbine forcing would follow directly from the local incident velocity and no such correction would be
required.

\subsection{Free atmosphere}\label{sec:free_atm}
Above the inversion, we set $f=\tau=0$, since the wind-farm forcing and the ABL turbulence do not directly affect the free atmosphere. As discussed in \S\ref{sec:governing_equation}, we also neglect the quadratic nonlinear terms in \eqref{2DBoussinesq}. Moreover, we assume the vertical variation of background velocity in the free atmosphere is negligible%
, so that $U'=0$. These simplifications reduces \eqref{2DBoussinesq} to
\begin{subequations}\label{FreeAtmos2}
\begin{align}
    \UH \px{u} &= - \px{p}, \label{fa1} \\
    \UH \px{w} &= - \pz{p}  + b, \label{fa2} \\
    \px{u} + \pz{w} &= 0,  \label{fa3} \\
    \UH \px{b} + N^2 w &= 0,  \label{fa4}
\end{align}
\end{subequations}
where $\UH=U(H)$. After eliminating $u$, $p$ and $b$ in favour of $w$ in \eqref{FreeAtmos2}, we obtain
\begin{equation}\label{weq_freeA}
    \pxx{w} + \pzz{w} = -({N^2}/{\UH^2}) \ w,
\end{equation}
Taking the Fourier transform of \eqref{weq_freeA} with respect to $x$ (denoted by hat), the well-known Helmholtz equation used in mountain-wave theory \citep{gill1982atmosphere} can be obtained:
\begin{equation}\label{wprime_eq}
    \pzz{\hat{w}(k,z)} + m^2 \hat{w}(k,z) = 0, \quad \text{with} \quad  m^2 = {N^2}/{\UH^2} - k^2,
\end{equation}
where $k$ is the horizontal wavenumber. The solution is
\begin{equation}\label{w_sol_raw}
    \hat{w}(k,z) =  \hat{w}^+(k,H)\, \e^{i m (z-H)}, \quad
    m = \begin{cases}
 i \sqrt{k^2 - {N^2}/{\UH^2}}, & \ \  \lvert k \rvert \geq N/\UH \\
 \sgn(k)\sqrt{{N^2}/{\UH^2} - k^2} , & \ \ \lvert k \rvert < N/\UH.
\end{cases}
\end{equation}
For $|k|\geq N/\UH$ the root is chosen to avoid exponential growth with height,
while for $|k| < N/\UH$ the branch giving upward group velocity is selected, so
that wave energy radiates away from the farm \citep{smith1980linear}. For completeness, the derivation of the latter is proved in Appendix~\ref{sec:rad_cond}. In the hydrostatic Boussinesq limit an analogous procedure gives $m=\sgn(k)N/\UH$. The above solution for the free atmosphere is similar to the inviscid solution for flow over mountains that has been extensively studied in mountain-wave theories both for the atmosphere and ocean \citep[see e.g.,][]{durran1990mountain,legg2021mixing}. Here, instead of a hill disturbing the stratified flow, the displacement of the capping inversion acts as the disturbance to the free atmosphere.

\subsection{Coupled dynamics at inversion layer}\label{sec:coupling_ABL_FA}
So far, the boundary-layer response and the free-atmospheric wave field have been derived separately, but neither problem is closed in isolation. The ABL requires an upper boundary condition at $z=H$, while the free-atmospheric solution depends on the unknown interfacial amplitude $\hat{w}^+(k,H)$. These two missing ingredients are supplied by coupling the two regions across the capping inversion through kinematic and dynamic conditions. The kinematic condition requires the inversion layer to move with the fluid
\begin{equation}\label{kin_cond}
{D\eta}/{Dt} = w^+ = w^-  \quad \rightarrow \quad
\hat w^+ = \hat{w}^-=i k \UH\hat\eta,
\end{equation}
where the second relation is the linearised Fourier representation for steady flow where $D \eta/D t=U_H \partial\eta/\partial x$. Note that the kinematic condition \eqref{kin_cond} is not explicitly imposed in the hydrostatic model of \cite{smith_gravity_2010} or in subsequent extensions based on the same vertically averaged ABL formulation \citep[e.g.][]{allaerts2019sensitivity}. As mentioned earlier, in these models, the vertical velocity is not dynamically coupled to the pressure and horizontal velocity fields within the boundary layer and therefore does not enter the interface coupling in the same manner as in the present formulation.
We next derive the dynamic condition by comparing the pressure distribution before and after displacing the capping inversion by $\eta$, which are denoted by $P$ (background pressure) and $P+p$ (background plus perturbation), respectively (figure \ref{fig:pressure-match}). Before displacement, the background pressures are in hydrostatic balance and all perturbation variables are zero: 
\begin{equation}\label{upper_layer}
    \pz{P} = -\rho^+ g /\rho_0 + N^2 (z-H) , 
\end{equation}
where $\rho^+$ is the density right above the inversion. Integrating \eqref{upper_layer} from $H$ to $H+\eta$ yields
\begin{equation}\label{before_disp}
    P(H+\eta) - P(H) = -\rho^+ g /\rho_0 \eta + N^2 \eta^2/2.
\end{equation}
After displacement (the right panel of figure \ref{fig:pressure-match}), we consider the vertical momentum equation to derive a relation between the total pressures, $P+p$, at $H$ and $H+\eta$ 
\begin{equation}
    \partial_z(P+p) =  -\rho^- g /\rho_0 - U \partial_x w 
\end{equation}
After the Taylor expansion of the last term around $z=H$ and integration, we find
\begin{equation}\label{after_disp}
    P(H+\eta) + p(H+\eta) - P(H) - p(H) = -\rho^- g /\rho_0 \eta -  U\partial_x w(H) \eta  + O(\eta^3)
\end{equation}
Subtracting \eqref{after_disp} from \eqref{before_disp} leads to 
\begin{equation}\label{p_match_dim}
    p(H) - p(H+\eta) = g_r \eta + N^2 \eta^2/2 + U\partial_x w(H) \eta + O(\eta^3)
\end{equation}
where $g_r= {(\rho^- - \rho^+)g}/{\rho_0} = {(\theta^+ - \theta^-)g}/{\theta_0}$ is the reduced gravity. Note that $p(H)$ is the pressure perturbation at the top of the ABL induced by the displacement of capping inversion, while $p(H+\eta)$ is the pressure of free-atmospheric waves at the interface. Hence, we can write $p^+ = p(H+\eta)$ and $p^-=p(H)$ following our preceding notation. 
\begin{figure}
\centering
\begin{tikzpicture}[>=Latex, thick, scale=0.8]

\tikzset{
  dot/.style={circle, draw=green!75!black, fill=white, line width=1pt, minimum size=2mm, inner sep=0pt},
  bluecurve/.style={blue!70!black, line width=1.2pt},
  redhatch/.style={pattern=north east lines, pattern color=cyan!75!black},
  myarrow/.style={-, black!70!black, line width=1.5pt}
}

\def\zH{0}
\def\etaZ{1.5}

\fill[redhatch] (0,-0.8) rectangle (4.5,\zH);

\draw[bluecurve] (0,\zH) -- (4.5,\zH);
\draw[bluecurve,dashed] (0,\etaZ) -- (4.5,\etaZ);

\node[dot] (LH) at (1.8,\zH-.1) {};
\node[dot] (Leta) at (1.8,\etaZ+.1) {};

\draw[decorate,decoration={brace,amplitude=6pt}]
(.5,\zH) -- (.5,\etaZ) node[midway,left=5pt] {$\eta$};

\node[anchor=east] at (-0.05,\etaZ) {$z=H+\eta$};
\node[anchor=east] at (-0.05,\zH) {$z=H$};

\node[anchor=west] at (1.3,\etaZ+0.4) {$P(H+\eta)$};
\node[anchor=west] at (1.3,\zH-0.5) {$P(H)$};

\node[green!50!black] at (4.3,0.45*\etaZ) {$\rho^+$};
\node[red!75!black] at (4.3,-0.35) {$\rho^-$};

\begin{scope}[xshift=6cm]

\draw[bluecurve]
(0,0)
.. controls (0.8,0) and (1.0,1.05*\etaZ) ..
(2.3,\etaZ)
.. controls (5.1,1.25*\etaZ) and (4.9,0) ..
(5.8,0);

\fill[redhatch]
(0,-0.8) --
(0,0)
.. controls (0.8,0) and (1.0,1.05*\etaZ) ..
(2.3,\etaZ)
.. controls (5.1,1.25*\etaZ) and (4.9,0) ..
(5.8,0) --
(5.8,-0.8) -- cycle;

\draw[bluecurve,dashed] (0.15,0) -- (5.8,0);

\node[dot] (Rt) at (2.7,\zH-.1) {};
\node[dot] (Rb) at (2.7,\etaZ+.1) {};

\node[anchor=west] at (2.,\etaZ+.4) {$P(H+\eta) + p(H+\eta)$};
\node[anchor=west] at (2.,-0.5) {$P(H) + p(H)$};

\draw[decorate,decoration={brace,amplitude=6pt}]
(2.3,\zH) -- (2.3,\etaZ) node[midway,left=5pt] {$\eta$};

\node[green!50!black] at (5.6,0.45*\etaZ) {$\rho^+$};
\node[red!75!black] at (5.6,-0.35) {$\rho^-$};

\end{scope}

\end{tikzpicture}
\caption{Schematic of pressures before (left) and after (right) displacing the interface.}
\label{fig:pressure-match}
\end{figure}
To identify the relative magnitude of terms in \eqref{p_match_dim}, we introduce the following scaling, obtained by requiring all retained terms in \eqref{2DBoussinesq} and \eqref{kin_cond} to be of comparable magnitude:
\begin{equation}
    (x,z):(L,H), \quad U: \UH, \quad  u: \epsilon \UH, \quad w = \epsilon \UH H/ L, \quad p: \epsilon \UH^2, \quad \eta : \epsilon H,
\end{equation}
where $L$ is the farm length, and $\epsilon$ is the ratio of perturbation to background velocities and assumed small. Defining $F_r = \UH/\sqrt{g_r H} $ and $F_N = \UH/(N H)$, we nondimensionalise \eqref{p_match_dim}
to obtain
\begin{equation}\label{p_match_nondim}
    p^- - p^+
    =
    F_r^{-2}\eta
    +\frac{1}{2}F_N^{-2}\epsilon \eta^2
    +\epsilon \frac{H}{L}
    U\partial_x w(H)\eta
    +O(\epsilon^2).
\end{equation}
For large wind farms, $H/L\lesssim 1$, so the third term on the right-hand side is higher order. If, in addition, $F_N^2/F_r^2 = g_r/N^2 H > \epsilon$, the first term dominates, reducing \eqref{p_match_dim} to
\begin{equation}\label{p_match_approx}
    p^- - p^+ \approx g_r \eta.
\end{equation}
Typically $g_r/N^2H=O(1)$, although this approximation becomes less accurate for ABLs with weak inversion layers. The approximate form of  \eqref{p_match_dim} (i.e.,  \eqref{p_match_approx}) has been widely used for interfaces above shallow-water layers \citep{smith_gravity_2010,allaerts2017,baines2022topographic}. %
An alternative derivation using pressure continuity at the capping inversion was also given by \cite{vosper2004inversion} in the context of linear mountain waves. Next we relate the quantities appearing in \eqref{p_match_approx} to $w$, since \eqref{realBLeq4w} is written in terms of the vertical velocity. For $k\neq0$, equations \eqref{fa1} and \eqref{fa3} give
\begin{equation}\label{p_above}
    \hat{p}^+ =
    \frac{U_H}{ik}\frac{\partial \hat{w}^+}{\partial z}.
\end{equation}
At the top of the ABL, we assume that $f$, $\tau$, and the nonlinear
perturbation terms are negligible, but retain the background shear immediately
below the inversion, denoted by $U_H'=U'(H^-)$. The horizontal momentum and
continuity equations then give
\begin{equation}\label{p_below}
    \hat{p}^- =
    \frac{1}{ik} \left(U_H\frac{\partial \hat{w}^-}{\partial z} - U_H'\hat{w}^- \right).
\end{equation}
Substituting Eqs.~\eqref{p_above} and \eqref{p_below} into
\eqref{p_match_approx}, using \eqref{w_sol_raw} to evaluate
$\partial\hat{w}^+/\partial z$, and expressing $\hat{\eta}$ in terms of
$\hat{w}$ through \eqref{kin_cond}, gives the boundary condition that closes
the boundary-layer problem,
\begin{equation}\label{key_BC}
    \pz{\hat{w}^-} = \frac{\Phi}{\UH^2} \, \hat{w}^-, \quad \text{where} \quad  \Phi = \begin{cases}
g_r + \UH \UH' + i\,\sgn(k)\, \UH \sqrt{N^2 - k^2 \UH^2}, & \lvert k \rvert < N/\UH \\
g_r + \UH \UH' - \UH\sqrt{k^2 \UH^2 - N^2}, & \lvert k \rvert > N/\UH.
\end{cases}
\end{equation}
In the hydrostatic limit, $\Phi=g_r+U_HU_H'+i\,\operatorname{sgn}(k)NU_H$. In our key boundary condition \eqref{key_BC}, the coefficient $\Phi$ encapsulates the effects of both interfacial waves, through the density jump $g_r$, and internal waves, through the background stratification $N$, on the ABL dynamics. As discussed by \cite{smith_gravity_2010}, this corresponds to the ABL pressure $p^-$ being the superposition of the free-atmospheric wave pressure $p^+$ and the interfacial-wave contribution $g_r\eta$ in \eqref{p_match_approx}. 

Although \eqref{key_BC} was derived independently within the present framework, in the limit $U_H'=0$ it reduces to the matching conditions used in the linear mountain-wave models of \cite{klemp1975dynamics} and \cite{vosper2004inversion}.
The lower-layer dynamics in these models differ substantially from the ABL model considered here, in particular through the absence of turbine forcing and turbulent stresses. At the top of the ABL, however, we assume that these effects, together
with the nonlinear perturbation terms, are negligible. Thus, when $U_H'=0$, the local equations governing the interface reduce to the corresponding mountain-wave equations, explaining the agreement between the matching conditions.

\subsection{Numerical implementation}\label{sec:numerics}

We employ a spectral discretisation in the horizontal direction, which provides a computationally efficient treatment of the governing equation. It also facilitates the implementation of our key vertical boundary condition \eqref{key_BC}, which is non-local and naturally expressed in Fourier space. It does, however, impose periodicity in $x$, requiring a domain sufficiently long for the farm wake to recover and the perturbations to vanish before reaching the boundaries.
Taking the Fourier transform of \eqref{realBLeq4w} in $x$ gives
\begin{equation}\label{L_N_f}
    \mathcal{L}\hat{w}
    =
    -\hat{\mathcal{N}}
    -\partial_z\hat{f},
\end{equation}
where, for $k\neq0$, $\mathcal{L}$ collects all terms that are linear in
$\hat{w}$:
\begin{equation}
    \mathcal{L} \hat{w}= U \left(\pzz{\hat{w}} - k^2 \hat{w} \right)
- U'' \hat{w}
- \frac{2}{ik}\pzz{}\left( \nu_t \pzz{\hat{w}} \right)
+ 2\nu_t ik \pzz{\hat{w}} \quad \text{for} \quad k \ne 0.
\end{equation}
For each horizontal wavenumber $k\neq0$, the Fourier-transformed governing equation is a fourth-order ordinary differential equation in $z$. The Fourier mode associated with $k=0$ requires separate treatment. For this mode, the streamwise-derivative term in the continuity equation \eqref{BLeq_linear_continuity} vanishes identically, leaving $\partial_z\hat{w}=0$, so that $\hat{w}(0,z)$ is independent of height. Together with the no-penetration condition at the surface, this gives
$\hat{w}(0,z) = 0$. 

The turbine forcing $f$ depends on the solution through the local velocity perturbation $u(x_n,z_h)$ and is therefore not known a priori. Moreover, the nonlinear term $\mathcal{N}(w)$ involves products of perturbation fields, whose Fourier transforms generate convolution sums and hence couple the different
horizontal wavenumbers. We therefore employ an iterative scheme in which both the turbine forcing and
the nonlinear terms on the right-hand side of \eqref{L_N_f} are evaluated using the solution from the previous iteration. Consequently, at each iteration, the right-hand side is known and \eqref{L_N_f} can be solved independently for each horizontal wavenumber $k$. The nonlinear terms are evaluated pseudo-spectrally, with products formed in physical space and dealiased using the two-thirds rule.

At each iteration, the inverse Fourier transform of $\hat{w}$ is already computed in order to evaluate the nonlinear terms in physical space. We therefore also recover the horizontal velocity directly in physical space by integrating the continuity equation \eqref{BLeq_linear_continuity},
\begin{equation}
    u(x,z)
    =
    -\int_{x_0}^x \pz{w(s,z)}\,\mathrm{d}s,
\end{equation}
with $x_0$ being the left computational boundary chosen sufficiently far upstream of the farm, where the perturbation velocity is taken to vanish. By periodicity, $u$ then also vanishes at the downstream boundary $x_0+L_x$, which is consistent only if the domain is long enough for the wake to fully recover, as discussed above. The flow field typically converges within 4--7 iterations.

In the vertical direction, we discretise the domain $z_0\leq z\leq H$ using second-order finite differences on a grid stretched towards the ground. The fourth-order linear operator then gives rise to a pentadiagonal matrix, allowing an efficient banded solver to be used independently for each horizontal
wavenumber. The four boundary conditions $\hat{w}=\partial_z\hat{w}=0$ at $z=z_0$ and $\partial_{zz}\hat{w}-(k^2+U''/U)\hat{w}=0$ and \eqref{key_BC} at $z=H$  are incorporated into the finite-difference discretisation using second-order one-sided differences. Once $w$ is obtained, the remaining fields $u$, $p$, and $\eta$ follow from \eqref{BLeq_linear_continuity},
\eqref{BLlin_zmom}, and \eqref{kin_cond}, respectively. The convergence of the solution with respect to the vertical and horizontal resolutions and the domain length is reported in \S\ref{sec:result}.

\section{Results}\label{sec:result}

\begin{figure}
    \centering
    \includegraphics[width=1\linewidth]{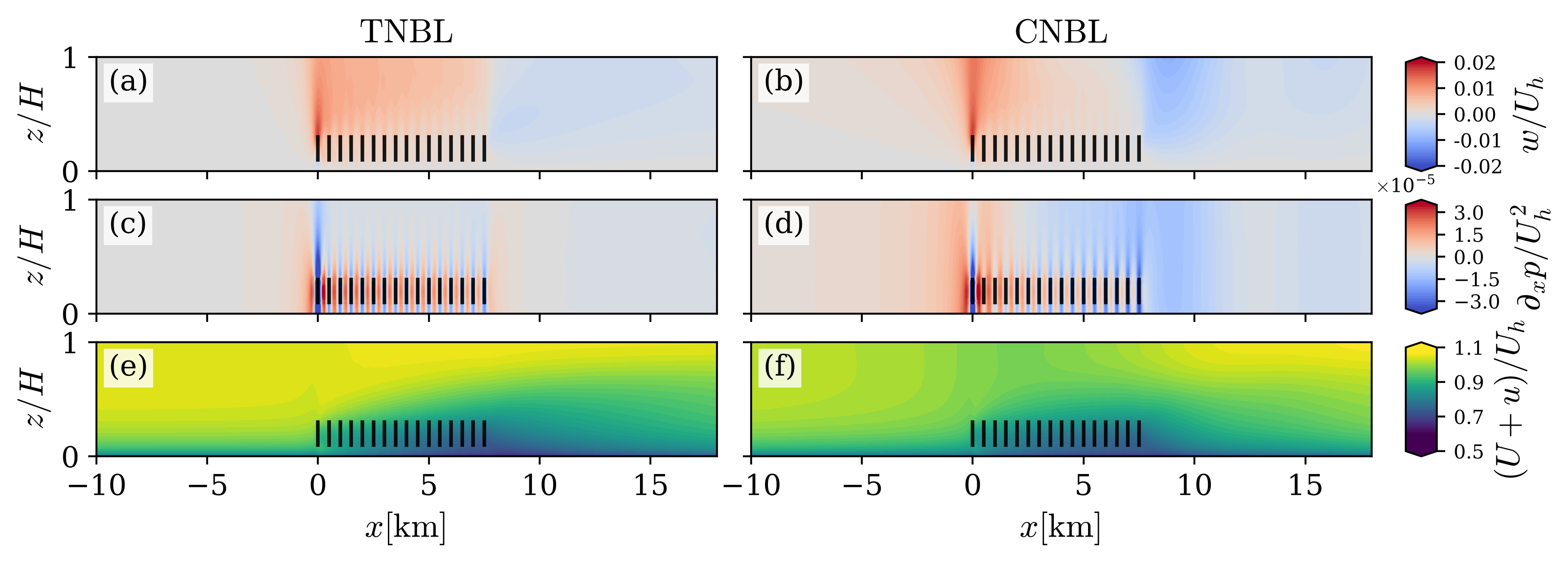}
   \caption{Normalised perturbations of vertical velocity (\textit{a,b}), streamwise pressure gradient (\textit{c,d}), and streamwise velocity (\textit{e,f}) for a TNBL against a CNBL. Turbine rows are shown by black lines.}
    \label{fig:qualitative}
\end{figure}

Figure~\ref{fig:qualitative} displays $w$, $\partial_x p$, and the total streamwise velocity $U+u$ for two boundary layers with $H=500$ m, $U_H=10\,\mathrm{m\,s^{-1}}$ and $z_0=10^{-4}\,\mathrm{m}$: (i) a truly neutral boundary layer (TNBL, $g_r=N=0$) with no gravity waves, and (ii) a conventionally neutral boundary layer (CNBL, $g_r=0.1\,\mathrm{m\,s}^{-2}, N=0.01\,\mathrm{s}^{-1}$, corresponding to $F_r=1.41$ and $F_N=2$) with the presence of internal and interfacial waves. In both scenarios, the semi-infinite wind farm consists of $N_r=$16 turbine rows with a constant thrust coefficient $C_T=0.75$, rotor diameter $D=100\,\mathrm{m}$, hub height $z_h=100\,\mathrm{m}$, and streamwise and spanwise spacings of $S_x=5D$ and $S_y=5D$, respectively. The domain is $500\,\mathrm{km}$ long to ensure the farm wake fully recovers, though only a small portion of it is shown in figure~\ref{fig:qualitative}.  It is discretised using $5120 \times 100$ grid points in the $x$- and $z$-directions, respectively, with vertical refinement near the surface. As discussed in \S \ref{sec:numerics}, a pseudo-spectral approach was used to model non-linear farm forcing and the advection terms. The solver was run for six iterations using an under-relaxation factor of $0.5$. Each iteration took less than $1\,\mathrm{s}$ on a standard laptop without code optimisation or parallelisation. The model relies on a single free parameter, $c_k$, discussed in  \S\ref{sec:farm_forcing_shorter}, whose value is kept fixed across all presented results in this paper.

As shown in figure~\ref{fig:qualitative}(a) for the TNBL case, flow deceleration within the wind farm creates an upward vertical displacement over the farm to satisfy continuity. This is followed by a downward displacement as the wake recovers downstream. However, in the CNBL case (figure~\ref{fig:qualitative}b), wind farm effects propagate much further upstream, and a significantly stronger downward motion is observed in the farm wake. The latter is driven by the restoring force of gravity from the overlying stably stratified capping inversion and free atmosphere, which force the flow downward in the lee side of the farm. Examining the pressure gradients, figure~\ref{fig:qualitative}(c) shows the localised pressure variations induced by the actuator disk forcing across each turbine row in the TNBL. The positive pressure gradients in front of each turbine row create an induction region (i.e., local blockage). In the CNBL (figure~\ref{fig:qualitative}d), these localised variations  are superimposed with a strong, large-scale pressure gradient generated by gravity waves. An adverse pressure gradient develops upstream of the farm, leading to strong global blockage effects. Within and downstream of the CNBL farm, a strong favourable pressure gradient forms, thereby accelerating wake recovery as shown in figure~\ref{fig:qualitative}f.

\begin{figure}
    \centering
    \includegraphics[width=0.62\linewidth]{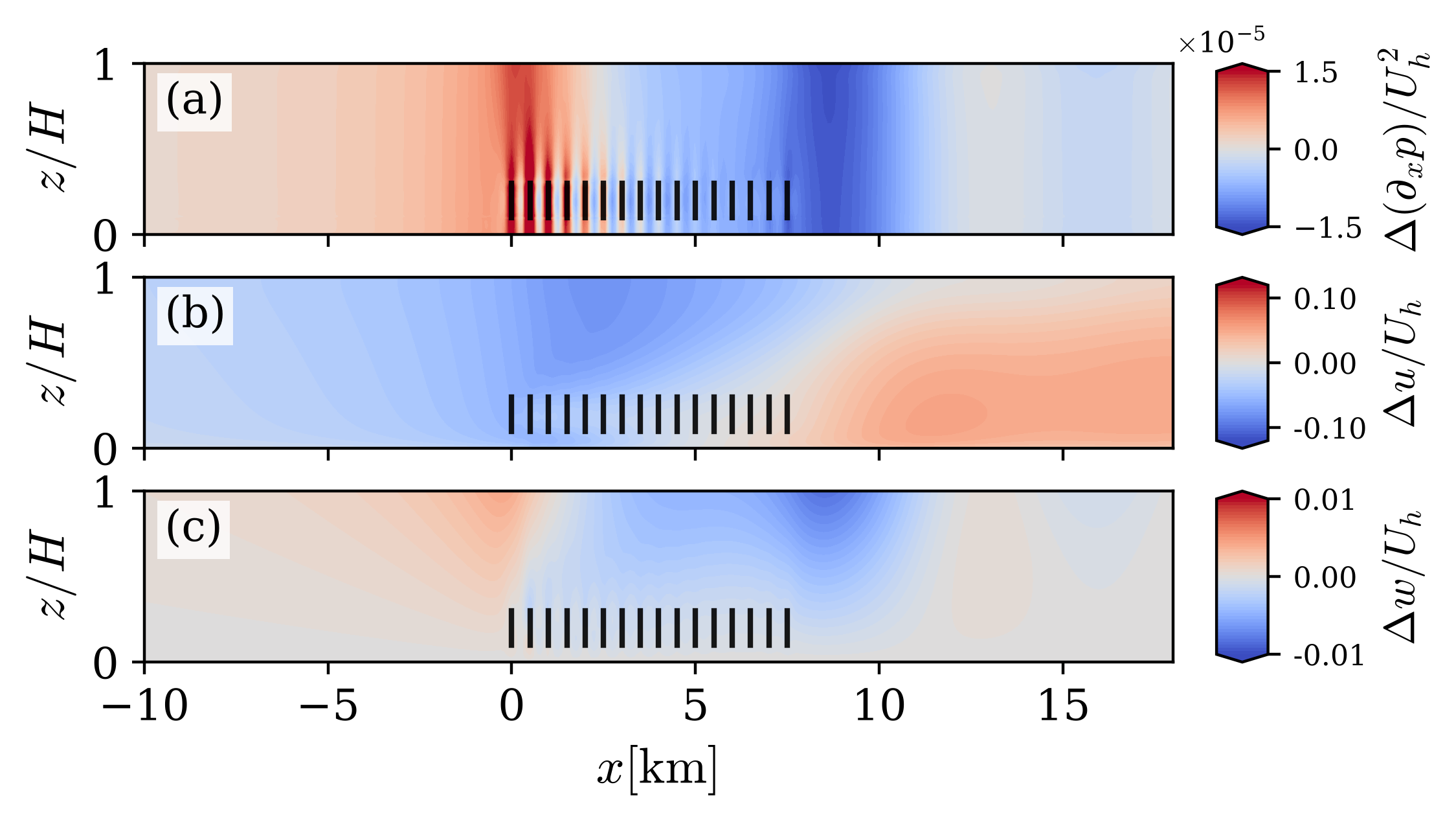}
   \caption{Difference between the CNBL and TNBL fields of figure~\ref{fig:qualitative}: (\textit{a}) streamwise pressure gradient, (\textit{b}) streamwise velocity, and (\textit{c}) vertical velocity, normalised as in figure~\ref{fig:qualitative}. Turbine rows are shown by black lines.}
    \label{fig:difference}
\end{figure}

Since the gravity-wave-induced perturbations are superimposed on the much stronger local variations around each turbine row, their footprint is difficult to discern in figure~\ref{fig:qualitative}. Figure~\ref{fig:difference} therefore shows the difference between the CNBL and TNBL fields. The large-scale pressure gradient (figure~\ref{fig:difference}a) is adverse over a region extending more than $10\,\mathrm{km}$ upstream of the farm and favourable within and downstream of it, with its strongest magnitude just behind the last row. The corresponding streamwise-velocity difference (figure~\ref{fig:difference}b) is negative ahead of and over the farm, where it represents the global blockage, and positive in the farm wake, where it represents the accelerated recovery; both extend over the full depth of the boundary layer, in contrast with the localised velocity deficit of the TNBL. The vertical-velocity difference (figure~\ref{fig:difference}c) shows the additional upward motion ahead of the farm and the strong downward motion over the farm wake associated with the restoring effect of the stratification.

 \begin{table}
    \centering
    \footnotesize %
    \begin{tabular}{l c c c c c c c c c c c}
        Case & $N \, [\mathrm{s}^{-1}]$ & $g_r \, [\mathrm{m\,s}^{-2}]$ & $F_r$ & $F_N$ & $z_0 \, [\mathrm{m}]$ & $H \, [\mathrm{m}]$ & $U_h \, [\mathrm{m\,s}^{-1}]$ & $U_H \, [\mathrm{m\,s}^{-1}]$ & $N_r$ & $S_y/D$ & $S_x/D$ \\
        \midrule
        JH & 0.006 & 0.022 & 3.20 & 2.5 & 0.1 & 1000 & 11.50 & 15 & 10 & 5.00 & 7.0 \\
        S1    & 0.006 & 0.034 & 2.06 & 2.00 & $2 \times 10^{-4}$ & 1000 & 10.96 & 12 & 20 & 5.33 & 7.5 \\
        Q00   & 0.006 & 0.170 & 0.92 & 2.00 & 0.1 & 1000 & 9.54  & 12 & 14 & 5.33 & 7.5 \\
        \bottomrule
    \end{tabular}
    \caption{\small Summary of LES datasets used for model validation. See  \citet{zhu_jhtdb-wind_2025} for more information on Case JH. Cases S1 and Q00 are discussed in \citet{allaerts2019sensitivity} and originally reported in \cite{allaerts2017,allaerts_gravity_2018}. Note $U_h=U(z=z_h)$. }
    \label{tab:simulation_parameters}
\end{table}

\begin{figure}
    \centering
    
    \begin{overpic}[width=0.6\textwidth]{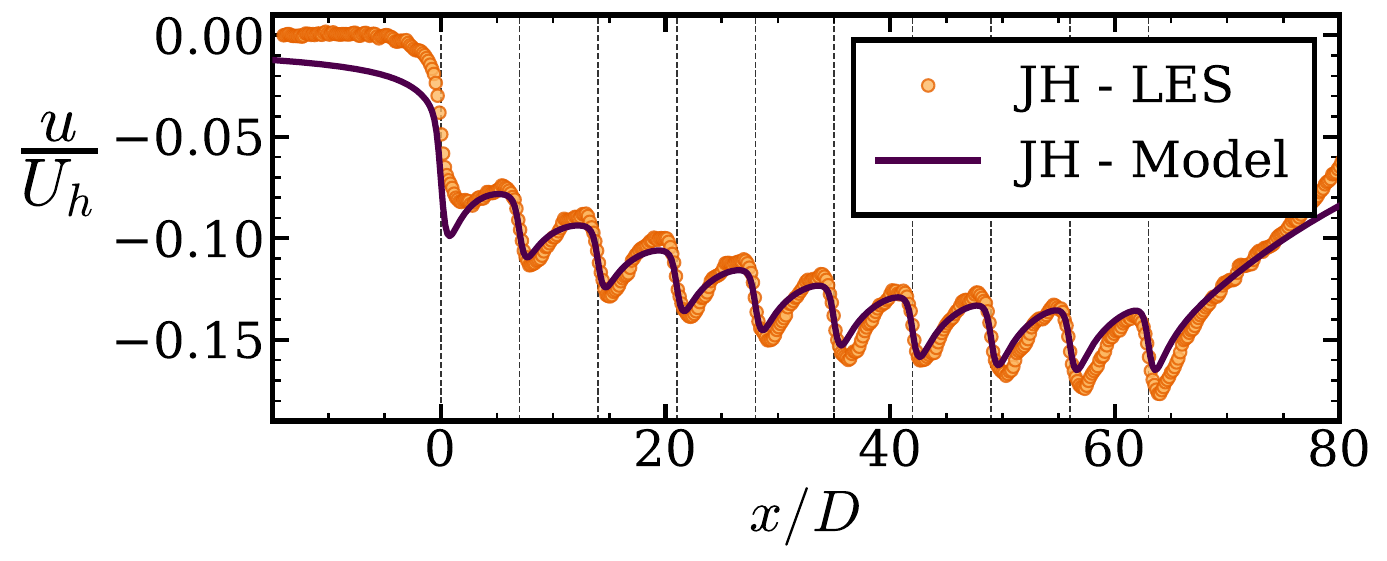}
        \put(-3, 40){(a)} 
    \end{overpic}
    
    \vspace{0.4cm} %
    
    \begin{overpic}[width=\textwidth]{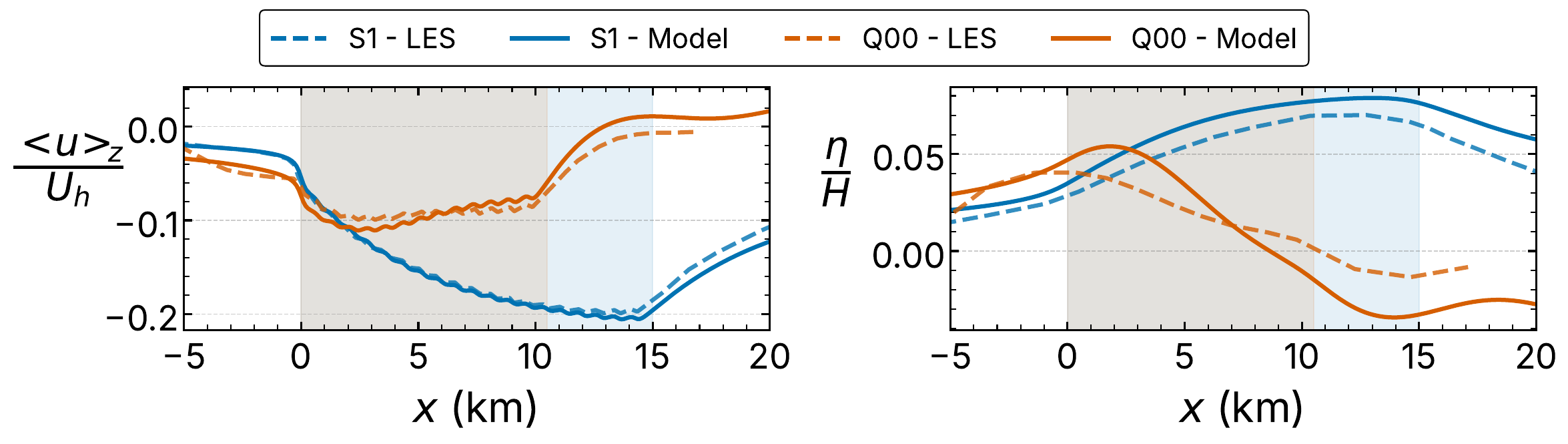}
        \put(12, 20){(b)}  %
        \put(61, 20){(c)}  %
    \end{overpic}
    
    \caption{(\textit{a}) Comparison of normalised velocity perturbation at hub height with the LES data of \cite{zhu_jhtdb-wind_2025}. (\textit{b}) Comparison of normalised velocity perturbation vertically averaged over $z=(0,2z_h)$ with the LES data reported in \cite{allaerts2019sensitivity}. (\textit{c}) Normalised capping inversion displacement $\eta/H$ comparing with the same dataset as in (\textit{b}). In (\textit{b}) and (\textit{c}), the shaded regions indicate the streamwise extents of the wind farms.}
    \label{fig:combined_validation}
\end{figure}

Overall, the above qualitative behaviours are consistent with the LES results reported in the literature \citep[e.g.,][]{lanzilao_parametric_2024, khan_dependence_2026}. In the following, we quantitatively compare model predictions with LES data of three semi-infinite wind farms. Some key atmospheric and farm properties for these three datasets are written in Table \ref{tab:simulation_parameters}. The inflow boundary layer is modelled using a logarithmic velocity profile within the surface layer ($z \leq z_s = 0.1H$), given by $U(z) = (u_*/\kappa) \ln(z/z_0)$, where $u_*$ is the friction velocity, and $\kappa=0.4$ is the von K\'arm\'an constant. For $z > z_s$, an empirical relation is employed that satisfies three conditions: it maintains a continuous value and slope with the logarithmic profile at $z = z_s$, and it asymptotes to $U_H$ as $z \to H$. This empirical profile is expressed as $U(z) = U_H - (U_H - U_s) \exp[-(z - z_s)/L_d]$, where $U_s$ is $U(z = z_s)$ computed from the logarithmic profile, and $L_d$ is the decay length scale of the empirical profile, defined as $L_d = (U_H - U_s)(\kappa/u_*) z_s$. %
Figure~\ref{fig:combined_validation}(a) shows streamwise variations of $u/U_h$ at
the hub height for both our reduced-order model and the LES data of
\cite{zhu_jhtdb-wind_2025}, which we refer to as the JH case in
Table~\ref{tab:simulation_parameters}. %
Within the farm the model reproduces the LES well, capturing both the row-by-row structure of the deficit
and its cumulative growth along the farm. Over the extent of the farm, the root-mean-square error of $u/U_h$ is $0.007$, i.e.\ $4\%$ of the maximum deficit, and the farm-averaged deficit is $0.127U_h$ in the model against $0.128U_h$ in the LES. Upstream of the farm, however,
the model predicts a deceleration that extends considerably further than that
seen in the LES. This difference is likely due to the streamwise extent
of the LES domain, which reaches only about $14D$ upstream of the first turbine row, whereas resolving the global blockage response requires domains extending much further ahead of the farm \citep{lanzilao_parametric_2024}. %

Next, we look at the LES data with much larger numerical domains reported in \cite{allaerts2019sensitivity}, focusing on cases denoted by S1 and Q00 in their paper. These two farms have different sizes (shaded areas in the figures \ref{fig:combined_validation}(b-c)) and different atmospheric conditions (see Table \ref{tab:simulation_parameters}). Case S1 has fairly weak stratification aloft, while Q00 is characterised by a strong potential temperature jump in the capping inversion. Based on Froude number $F_r$, S1 is supercritical and Q00 is subcritical as shown in Table \ref{tab:simulation_parameters}. In the subcritical case, trapped interfacial waves in the capping inversion layer can travel upstream, creating strong upwind pressure perturbations and thereby blockage \citep{smith_gravity_2010}. Figure~\ref{fig:combined_validation}(b) compares the streamwise velocity
perturbations averaged vertically over $0 \leq z \leq 2z_h$, denoted by
$\langle \cdot \rangle_z$, which reflects the format in which the LES data
were reported by \citet{allaerts2019sensitivity}. Overall, the model shows very good agreement with the LES data given the simplicity of the reduced-order framework: over the extent of the farms, the root-mean-square errors of $\langle u\rangle_z/U_h$ are $0.004$ for S1 and $0.009$ for Q00 ($2\%$ and $9\%$ of the respective maximum deficits), and the farm-averaged deficits are $0.164U_h$ (S1) and $0.093U_h$ (Q00) in the model against $0.162U_h$ and $0.090U_h$ in the LES. While Q00 experiences a considerably stronger blockage in the upwind region, its wake recovers faster than S1 due to the strong favourable pressure gradient created by gravity waves downstream as qualitatively discussed earlier in figure \ref{fig:qualitative}. In the strongly stratified Q00 case, the model overpredicts farm-wake recovery, resulting in a slightly positive $u$. This indicates an overestimation of the gravity-wave-induced favourable pressure gradient, potentially due to the model's 2D simplification. Finally, Figure~\ref{fig:combined_validation}(c) shows the vertical displacement of the capping inversion, $\eta$, normalised by the ABL thickness, $H$. Consistent with the LES data, the model indicates that the maximum displacement occurs near the end of the farm for the supercritical case (S1), whereas it shifts to the beginning of the farm for the subcritical case (Q00).

All results presented in this section retain the non-linear advection terms. Repeating the three LES cases with these terms omitted, so that only the non-linearity of the turbine forcing is retained, changes the maximum velocity deficit by $0.6$--$2.3\%$ and the maximum inversion displacement by $0.2$--$4.4\%$, confirming that the laterally averaged perturbations are small enough for the linearisations to be acceptable here. However, the non-linear terms are expected to be more important in a three-dimensional formulation, in which the local velocity deficits in the near wakes of the turbines are considerably larger than their lateral averages.

The sensitivity of the results to the numerical parameters was also assessed for the Q00 and JH cases, in terms of the relative $L_2$ difference of the hub-height velocity over the streamwise extent of the farm. Increasing the number of vertical grid points from the value used here ($N_z=100$) to $N_z=400$ changes it by less than $2\%$, refining the grid in the streamwise direction by a factor of four changes it by less than $0.5\%$, and increasing the domain length from $1000$ to $4000\,\mathrm{km}$ for the subcritical case Q00, which is the most demanding in this respect because its interfacial waves propagate upstream, changes it by less than $0.1\%$. The results of figures~\ref{fig:qualitative}--\ref{fig:combined_validation} are therefore converged to within a few per cent.

\section{Conclusions and outlook}

This work developed a reduced-order numerical model for the two-way coupling
between turbulent wakes and atmospheric gravity waves in large-scale wind farms.
The starting point is the steady Boussinesq system, retained in
non-hydrostatic form in both the boundary layer and the free atmosphere, so that the horizontal velocity, vertical velocity and pressure remain dynamically coupled throughout.
Within the ABL, eliminating the pressure and horizontal velocity reduces the
system to a single equation for the vertical velocity, which retains the
continuous vertical structure of the boundary layer. Above the inversion, the same system reduces to a Helmholtz equation whose solution is selected by the radiation condition, giving
an analytical description of the wave field aloft.

The two regions are joined at the capping inversion through kinematic and
dynamic conditions. The former requires continuity of $w$ across the inversion,
while the latter, obtained from the pressure jump associated with the displaced
interface, yields a Robin boundary condition \eqref{key_BC} for the vertical
velocity at the top of the ABL. The gravity-wave feedback on the boundary layer
is therefore expressed through a single upper boundary condition. The boundary-layer problem is thereby closed without an iterative exchange between separate micro- and meso-scale components.

The turbine forcing enters the governing equations as a momentum sink acting on
the resolved flow field, obtained by laterally averaging a three-dimensional
actuator-disc representation of the farm. Wake recovery and local blockage
therefore develop as part of the solution rather than being prescribed by an
engineering wake model. The nonlinear dependence of the thrust on the local
velocity is retained, as are the nonlinear advection terms in the governing
equations. The resulting system is solved with a mixed pseudo-spectral and
finite-difference discretisation in which the wavenumbers decouple, and the
forcing and nonlinear terms are treated as source terms updated iteratively. The solution typically converges within a handful of iterations, so that a complete case is obtained in seconds on a standard laptop, several orders of magnitude below the cost of a high fidelity simulation. Aside from the tuning
constant $c_k$ entering the layout correction, which is specific to the laterally averaged formulation and has been calibrated against the LES datasets used here, the model introduces no free
parameters.

Comparisons with three LES datasets spanning different atmospheric regimes show good agreement in both the internal wind-farm flow and the large-scale
gravity-wave response. The model reproduces the row-by-row structure of the
velocity deficit within the farm, the upstream blockage associated with the
adverse pressure gradient ahead of the farm, and the accelerated wake recovery driven by the favourable pressure gradient within and downstream of it. It also captures the shift in the location of maximum inversion displacement between supercritical and subcritical conditions. These comparisons involve a single calibrated constant and three datasets, and should be regarded as a first assessment of the framework rather than a comprehensive validation. The combination of physics-based wake and gravity-wave coupling in one single framework with a computational cost far below that of high-fidelity simulation makes this approach a promising candidate for applications that require many evaluations, such as parameter sweeps over atmospheric conditions. Subject to the extension to finite farms and to validation over a wider range of conditions, it could also be used to assess farm-to-farm interactions. Moreover, since the gravity-wave response of the free atmosphere is obtained analytically rather than being numerically resolved, the approach could provide a fast complement to the wind-farm parametrisations of mesoscale models, in which the farm is represented as a drag within the resolved boundary layer and the gravity-wave feedback is not explicitly represented.

Several simplifications limit the present formulation and indicate where it can usefully be extended. The farm is taken to be semi-infinite in the lateral direction, so that the flow is solved in two dimensions. While lateral heterogeneity within the farm is empirically accounted for  through the layout correction $\beta_n$, effects such as flow redirection around the farm sides and the associated three-dimensional wave field are absent. The eddy viscosity is evaluated from the background shear alone, and may therefore under-represent the enhanced mixing generated within densely spaced farms. The Coriolis force and boundary-layer stratification are not included, so that wind veer and the supergeostrophic jet near the inversion are not represented. The periodic streamwise boundary conditions require long domains, which could be avoided in future by a fringe-region treatment. Extending the framework to finite farms, validating it against a broader set of high-fidelity data and comparing it with existing layer-averaged models, and incorporating these effects are the natural next steps. 

\appendix

\section{Time-dependent free-atmospheric waves and radiation condition}\label{sec:rad_cond} 

The steady equations alone do not determine the appropriate branch of the square root appearing in the vertical wavenumber. This ambiguity is resolved by considering the time-dependent problem and imposing the radiation condition. To this end, we retain the time derivatives in the free-atmospheric equations \eqref{FreeAtmos2}, yielding
\begin{subequations}\label{FreeAtmos_dt}
\begin{align}
\pt{u} +  \UH \px{u} &= - \px{p}, \label{fat1} \\
\pt{w} +  \UH \px{w} &= - \pz{p}  + b,  \label{fat2} \\
\px{u} + \pz{w} &= 0, \label{fat3} \\
\pt{b} + \UH \px{b} + N^2 w &= 0. \label{fat4}
\end{align}
\end{subequations}
We first use \eqref{fat3} to substitute for $u'$ in \eqref{fat1} and then seek plane-wave solutions of the form
\begin{equation}
w = \tilde{w} \e^{i(k x + m z -\omega t)}, \quad
p = \tilde{p} \e^{i(k x + m z -\omega t)}, \quad
b = \tilde{b} \e^{i(k x + m z -\omega t)} ,
\end{equation}
which, upon substitution into \eqref{fat1}, \eqref{fat2}, and \eqref{fat4}, lead to the linear system
\begin{equation}
   \begin{bmatrix}
	i \omega (m/k) - i m \UH	    &  ik          	& 0    \\
	-i \omega + i k \UH	            	& i m        	& -1    \\
	   N^2  & 0     & -i \omega + i k \UH	
	\end{bmatrix} 
	\begin{bmatrix} \tilde{w} \\ \tilde{p} \\ \tilde{b} \end{bmatrix}
        = \boldsymbol{0}. 
\end{equation}
Requiring a non-trivial solution yields the dispersion relation
\begin{equation}\label{raw_disprel}
(\omega - k \UH)^2 = \frac{k^2 N^2}{k^2 + m^2},
\end{equation}
which can be written as
\begin{equation}\label{disprel}
\omega = k \UH + \omega_i, \qquad
\omega_i = \pm \frac{|k| N}{\sqrt{k^2 + m^2}},
\end{equation}
where $\omega_i$ is the intrinsic frequency, and its sign remains to be determined. Squaring both sides of \eqref{disprel} and then differentiating with respect to $m$ give
\begin{equation}\label{pos_gvel}
2 \omega_i \frac{\partial \omega_i}{\partial m}
= - \frac{2 m k^2 N^2}{(k^2 + m^2)^2}.
\end{equation}
The radiation condition requires the vertical group velocity to be directed upward, i.e.  
\begin{equation}\label{rad_condit}
    \frac{\partial \omega }{\partial m} = \frac{\partial \omega_i }{\partial m}>0
\end{equation}
ensuring that wave energy propagates away from the boundary and into the free atmosphere. Equations \eqref{rad_condit} and \eqref{pos_gvel} therefore imply
\begin{equation}\label{cond_x1}
\sgn(m) = - \sgn(\omega_i).
\end{equation}
For steady waves $\omega = 0$, we have $\omega_i = - k \UH$ and hence $\sgn(k) = -\sgn(\omega_i) $ since the background velocity satisfies $\UH>0$. Combining this result with \eqref{cond_x1} yields
\begin{equation}\label{sgn_mk}
\sgn(k) = \sgn(m).
\end{equation}
Thus, upward energy propagation requires the vertical and horizontal wavenumbers to have the same sign. This is the branch used throughout the paper. Using $\omega = 0$ in the dispersion relation \eqref{disprel} together with \eqref{sgn_mk} yields
\begin{equation}
m = \sgn(k)\sqrt{\frac{N^2}{\UH^2} - k^2}.
\end{equation}

\begin{Backmatter}

\paragraph{Acknowledgements}
The authors used ChatGPT (OpenAI; GPT-5.6 Sol) for assistance with reviewing algebraic operations, and language editing during the preparation of the manuscript. 

\paragraph{Funding Statement}
M.B and H.A.K are supported by the Engineering and Physical Sciences Research Council's impact acceleration account (IAA), grant EP/X525546/1. H.A.K is also supported by the Engineering and Physical Sciences Research Council, grant EP/Y021479/1. 

\paragraph{Declaration of Interests}
The authors declare no conflict of interest.

\paragraph{Author Contributions}
H.K. and M.B. contributed equally to this research, including conceptualisation, model development, code implementation, manuscript writing, and critical review. The order of authorship was determined by a coin toss.

\paragraph{Data Availability Statement}
Raw data are available from the corresponding authors.

\paragraph{Ethical Standards}
The research meets all ethical guidelines, including
adherence to the legal requirements of the study country.

\bibliographystyle{plainnat}
\bibliography{wind_gfd,jfm}

\begin{thebibliography}{28}
\providecommand{\natexlab}[1]{#1}
\providecommand{\url}[1]{\texttt{#1}}
\expandafter\ifx\csname urlstyle\endcsname\relax
  \providecommand{\doi}[1]{doi: #1}\else
  \providecommand{\doi}{doi: \begingroup \urlstyle{rm}\Url}\fi

\bibitem[Allaerts and Meyers(2017)]{allaerts2017}
D.~Allaerts and J.~Meyers.
\newblock Boundary-layer development and gravity waves in conventionally
  neutral wind farms.
\newblock \emph{Journal of Fluid Mechanics}, 814:\penalty0 95--130, 2017.
\newblock \doi{https://doi.org/10.1017/jfm.2017.11}.

\bibitem[Allaerts and Meyers(2018)]{allaerts_gravity_2018}
Dries Allaerts and Johan Meyers.
\newblock Gravity {Waves} and {Wind}-{Farm} {Efficiency} in {Neutral} and
  {Stable} {Conditions}.
\newblock \emph{Boundary-Layer Meteorology}, 166\penalty0 (2):\penalty0
  269--299, February 2018.
\newblock ISSN 0006-8314, 1573-1472.
\newblock URL \url{http://link.springer.com/10.1007/s10546-017-0307-5}.

\bibitem[Allaerts and Meyers(2019)]{allaerts2019sensitivity}
Dries Allaerts and Johan Meyers.
\newblock Sensitivity and feedback of wind-farm-induced gravity waves.
\newblock \emph{Journal of Fluid Mechanics}, 862:\penalty0 990--1028, 2019.

\bibitem[Baines(2022)]{baines2022topographic}
Peter~G Baines.
\newblock \emph{Topographic effects in stratified flows}.
\newblock Cambridge University Press, 2022.

\bibitem[Bastankhah and Port{\'e}-Agel(2014)]{bastankhah2014new}
M~Bastankhah and F~Port{\'e}-Agel.
\newblock A new analytical model for wind-turbine wakes.
\newblock \emph{Renew. energy}, 70:\penalty0 116--123, 2014.

\bibitem[Bastankhah et~al.(2024)Bastankhah, Mohammadi, Lees, Diaz, Buxton, and
  Ivanell]{bastankhah2024farmwake}
Majid Bastankhah, Mohammad~Mehdi Mohammadi, Charlie Lees, Gonzalo Pablo~Navarro
  Diaz, Oliver~RH Buxton, and Stefan Ivanell.
\newblock A fast-running physics-based wake model for a semi-infinite wind
  farm.
\newblock \emph{Journal of Fluid Mechanics}, 985:\penalty0 A43, 2024.

\bibitem[Belcher et~al.(2003)Belcher, Jerram, and Hunt]{belcher-Hunt2003}
SE~Belcher, N~Jerram, and JCR Hunt.
\newblock Adjustment of a turbulent boundary layer to a canopy of roughness
  elements.
\newblock \emph{Journal of Fluid Mechanics}, 488:\penalty0 369--398, 2003.

\bibitem[Devesse et~al.(2024)Devesse, Lanzilao, and
  Meyers]{devesse_mesomicro_2024}
Koen Devesse, Luca Lanzilao, and Johan Meyers.
\newblock A meso–micro atmospheric perturbation model for wind farm blockage.
\newblock \emph{Journal of Fluid Mechanics}, 998:\penalty0 A63, November 2024.
\newblock ISSN 0022-1120, 1469-7645.
\newblock \doi{10.1017/jfm.2024.868}.
\newblock URL
  \url{https://www.cambridge.org/core/product/identifier/S0022112024008681/type/journal_article}.

\bibitem[Durran(1990)]{durran1990mountain}
Dale~R Durran.
\newblock Mountain waves and downslope winds.
\newblock In \emph{Atmospheric processes over complex terrain}, pages 59--81.
  Springer, 1990.

\bibitem[Ebenhoch et~al.(2017)Ebenhoch, Muro, Dahlberg, Berkesten~H{\"a}gglund,
  and Segalini]{ebenhoch2017}
R.~Ebenhoch, B.~Muro, J.{\AA}. Dahlberg, P.~Berkesten~H{\"a}gglund, and
  A.~Segalini.
\newblock A linearized numerical model of wind-farm flows.
\newblock \emph{Wind Energy}, 20\penalty0 (5):\penalty0 859--875, 2017.

\bibitem[Gill(2016)]{gill1982atmosphere}
Adrian~E Gill.
\newblock \emph{Atmosphere---ocean dynamics}.
\newblock Elsevier, 2016.

\bibitem[Khan et~al.(2025)Khan, Allaerts, Watson, and
  Churchfield]{khan_investigating_2025}
Mehtab~Ahmed Khan, Dries Allaerts, Simon~J. Watson, and Matthew~J. Churchfield.
\newblock Investigating the relationship between simulation parameters and flow
  variables in simulating atmospheric gravity waves for wind energy
  applications.
\newblock \emph{Wind Energy Science}, 10\penalty0 (6):\penalty0 1167--1185,
  July 2025.
\newblock ISSN 2366-7443.

\bibitem[Khan et~al.(2026)Khan, Churchfield, and Watson]{khan_dependence_2026}
Mehtab~Ahmed Khan, Matthew~J. Churchfield, and Simon~J. Watson.
\newblock Dependence of wind-farm-induced gravity waves and wind farm
  performance on non-dimensional atmospheric parameters and simulation
  configuration.
\newblock \emph{Wind Energy Science}, 11\penalty0 (5):\penalty0 1631--1652, May
  2026.
\newblock ISSN 2366-7443.
\newblock \doi{10.5194/wes-11-1631-2026}.
\newblock URL \url{https://wes.copernicus.org/articles/11/1631/2026/}.

\bibitem[Klemp and Lilly(1975)]{klemp1975dynamics}
Joseph~B Klemp and DR~Lilly.
\newblock The dynamics of wave-induced downslope winds.
\newblock \emph{Journal of Atmospheric Sciences}, 32\penalty0 (2):\penalty0
  320--339, 1975.

\bibitem[Lanzilao and Meyers(2024)]{lanzilao_parametric_2024}
L.~Lanzilao and J.~Meyers.
\newblock A parametric large-eddy simulation study of wind-farm blockage and
  gravity waves in conventionally neutral boundary layers.
\newblock \emph{Journal of Fluid Mechanics}, 979:\penalty0 A54, January 2024.
\newblock ISSN 0022-1120, 1469-7645.
\newblock \doi{10.1017/jfm.2023.1088}.

\bibitem[Lanzilao and Meyers(2025)]{lanzilao_wind-farm_2025}
Luca Lanzilao and Johan Meyers.
\newblock Wind-farm wake recovery mechanisms in conventionally neutral boundary
  layers.
\newblock \emph{Journal of Fluid Mechanics}, 1015:\penalty0 A5, July 2025.
\newblock ISSN 0022-1120, 1469-7645.
\newblock \doi{10.1017/jfm.2025.10320}.

\bibitem[Legg(2021)]{legg2021mixing}
Sonya Legg.
\newblock Mixing by oceanic lee waves.
\newblock \emph{Annual Review of Fluid Mechanics}, 53\penalty0 (1):\penalty0
  173--201, 2021.

\bibitem[Maas(2023)]{maas_gigawatt_2023}
Oliver Maas.
\newblock From gigawatt to multi-gigawatt wind farms: wake effects, energy
  budgets and inertial gravity waves investigated by large-eddy simulations.
\newblock \emph{Wind Energy Science}, 8\penalty0 (4):\penalty0 535--556, April
  2023.
\newblock ISSN 2366-7443.

\bibitem[Ott(2011)]{ott2011Fuga}
S.~Ott.
\newblock Linearised cfd models for wakes.
\newblock Technical report, Danmarks Tekniske Universitet, Ris{\o}
  Nationallaboratoriet for B{\ae}redygtig Energi, 2011.

\bibitem[Segalini(2017)]{segalini_linearized_2017}
Antonio Segalini.
\newblock Linearized simulation of flow over wind farms and complex terrains.
\newblock \emph{Philosophical Transactions of the Royal Society A:
  Mathematical, Physical and Engineering Sciences}, 375\penalty0
  (2091):\penalty0 20160099, March 2017.
\newblock ISSN 1364-503X.
\newblock \doi{10.1098/rsta.2016.0099}.
\newblock URL \url{https://doi.org/10.1098/rsta.2016.0099}.

\bibitem[Segalini(2021)]{segalini_analytical_2021}
Antonio Segalini.
\newblock An analytical model of wind-farm blockage.
\newblock \emph{Journal of Renewable and Sustainable Energy}, 13\penalty0 (3),
  2021.

\bibitem[Smith(1980)]{smith1980linear}
Ronald~B Smith.
\newblock Linear theory of stratified hydrostatic flow past an isolated
  mountain.
\newblock \emph{Tellus}, 32\penalty0 (4):\penalty0 348--364, 1980.

\bibitem[Smith(2010)]{smith_gravity_2010}
Ronald~B Smith.
\newblock Gravity wave effects on wind farm efficiency.
\newblock \emph{Wind Energy}, 13\penalty0 (5):\penalty0 449--458, 2010.

\bibitem[Stipa et~al.(2024{\natexlab{a}})Stipa, Ajay, Allaerts, and
  Brinkerhoff]{stipa2024tosca}
Sebastiano Stipa, Arjun Ajay, Dries Allaerts, and Joshua Brinkerhoff.
\newblock Tosca--an open-source, finite-volume, large-eddy simulation (les)
  environment for wind farm flows.
\newblock \emph{Wind Energy Science}, 9\penalty0 (2):\penalty0 297--320,
  2024{\natexlab{a}}.

\bibitem[Stipa et~al.(2024{\natexlab{b}})Stipa, Ajay, Allaerts, and
  Brinkerhoff]{stipa_multi-scale_2024}
Sebastiano Stipa, Arjun Ajay, Dries Allaerts, and Joshua Brinkerhoff.
\newblock The multi-scale coupled model: a new framework capturing wind
  farm–atmosphere interaction and global blockage effects.
\newblock \emph{Wind Energy Science}, 9\penalty0 (5):\penalty0 1123--1152, May
  2024{\natexlab{b}}.
\newblock ISSN 2366-7443.
\newblock \doi{10.5194/wes-9-1123-2024}.
\newblock URL \url{https://wes.copernicus.org/articles/9/1123/2024/}.

\bibitem[Vallis(2017)]{vallis2017atmospheric}
Geoffrey~K Vallis.
\newblock \emph{Atmospheric and oceanic fluid dynamics}.
\newblock Cambridge University Press, 2017.

\bibitem[Vosper(2004)]{vosper2004inversion}
Simon~B Vosper.
\newblock Inversion effects on mountain lee waves.
\newblock \emph{Quarterly Journal of the Royal Meteorological Society},
  130\penalty0 (600):\penalty0 1723--1748, 2004.

\bibitem[Zhu et~al.(2025)Zhu, Xiao, Narasimhan, Martinez-Tossas, Schnaubelt,
  Lemson, Yao, Szalay, Gayme, and Meneveau]{zhu_jhtdb-wind_2025}
Xiaowei Zhu, Shuolin Xiao, Ghanesh Narasimhan, Luis~A. Martinez-Tossas, Michael
  Schnaubelt, Gerard Lemson, Hanxun Yao, Alexander~S. Szalay, Dennice~F. Gayme,
  and Charles Meneveau.
\newblock {JHTDB}-wind: a web-accessible large-eddy simulation database of a
  wind farm with virtual sensor querying.
\newblock \emph{Wind Energy Science}, 10\penalty0 (12):\penalty0 2821--2840,
  December 2025.
\newblock ISSN 2366-7443.

\end{thebibliography}

\end{Backmatter}

\end{document}